\documentclass[pra,twocolumn,superscriptaddress,showpacs]{revtex4}
\usepackage{amsmath,graphicx}
\newcommand{\ket}[1]{\vert #1 \rangle}
\newcommand{\bra}[1]{\langle #1 \vert}
\begin{document}
\title{Demonstration of a programmable source of two-photon
multipartite entangled states}
\author{Simone Cialdi}
\affiliation{Dipartimento di Fisica dell'Universit\`a degli Studi
di Milano, I-20133 Milano, Italia.}
\affiliation{INFN, Sezione di Milano, I-20133 Milano, Italia.}
\author{Davide Brivio}
\affiliation{Dipartimento di Fisica dell'Universit\`a degli Studi
di Milano, I-20133 Milano, Italia.}
\author{Matteo G.A.~Paris}
\affiliation{Dipartimento di Fisica dell'Universit\`a degli Studi
di Milano, I-20133 Milano, Italia.}
\affiliation{CNISM UdR Milano Universit\`a, I-20133, Milano, Italia.}
\date{\today}
\begin{abstract}
We suggest and demonstrate a novel source of two-photon multipartite
entangled states which exploits the transverse spatial structure of
spontaneous parametric downconversion together with a programmable
spatial light modulator (SLM). The $1$D SLM is used to perform
polarization entanglement purification and to realize arbitrary
phase-gates between polarization and momentum degrees of freedom of
photons. We experimentally demonstrate our scheme by generating 
two-photon three qubit linear cluster states with high fidelity 
using a diode laser pump with a limited coherence time and power on 
the crystal as low as $\sim 2.5$mW.
\end{abstract}
\maketitle
\section{Introduction}
Multiqubit entangled states, e.g. cluster states, are key resources
to realize several protocols of quantum information processing, including
measurement based quantum computation \cite{brg01,nie06,val08}, quantum
communication \cite{cle99} and quantum error correction \cite{dirk01}.
Besides, they found applications in advanced fundamental tests of quantum
nonlocality \cite{mer90,cabs,expm1,expm2}.
Basically, there are two ways to generate multiqubit entangled states,
e.g. cluster states. On one hand, one may increase the number of entangled
photons \cite{zha03,wal05,kie05,pre07}. On the other hand one may use
different degrees of freedom of the same pair of photons 
\cite{chen07,val07,val08,nev09} achieving so-called hyperentanglement.
The second method offers a larger robustness against decoherence and 
nonunit detector efficiency. Four and six multiphoton cluster states 
have been experimentally created \cite{zha03,wal05,kie05} as well as two-photon four- 
\cite{val07,chen07,val08,tok08,mai01,cin05,bar05,barr05,sch06,par07,lan09,val09}
and six-qubit cluster states \cite{cec09}.
\par
In this paper we suggest and demonstrate a novel scheme to generate
two-photon multipartite entangled states which exploits the transverse
spatial structure of spontaneous parametric downconversion and a
programmable spatial light modulator ($1$D SLM) based on a liquid
crystal display.  This kind of devices have been already used as
pulse shaper  for Bell state generation \cite{day07}, as 
amplitude modulators for momentum imaging and qudit generation
\cite{lima09} as well as diffractive elements to operate on orbital
angular momentum \cite{yao06}. Here we employ SLM in an innovative way
to realize two-photon multiqubit/qudit entangled states and demonstrate
its use in the generation of two-photon three-qubit linear cluster
states with high fidelity. 
\par
The novelty of our setup is twofold.  
On the one hand, we use the SLM for purification, and this allows us 
to dramatically decrease the spectral and angular filtering of 
downconverted  photons, which is the method generally used to prevent 
the degradation of the purity.
Moreover the SLM may be externally controlled, via software, and this
makes our method more easily adjustable for the different implementations, 
compared to purification schemes that involve the use of suitably prepared
crystals along the path of the downconverted photons \cite{kw09}.  On
the other hand, we fully exploit the properties of the SLM to realize
arbitrary phase-gates between polarization and momentum degrees of
freedom.  In this way, we obtain an effective, low cost, source of
two-photon multipartite entanglement using a pump with low power and a
limited coherence time.
\par
The paper is structured as follows: In the next Section we describe
our PDC system in some details and illustrate the purification method
based on the use of SLM. In Section III we address the generation of 
two-photon multiqubit/qudit entangled states and describe our
experimental setup, used to demonstrate the generation of two-photon 
three-qubit linear cluster states with high fidelity. 
Section IV closes the paper with some concluding remarks.
\section{Polarization entanglement and purification}
The
first step in our scheme is the generation of polarization entangled
states by spontaneous parametric down conversion (SPDC) in two adjacent
BBO crystals oriented with their optical axes aligned in perpendicular
planes \cite{Kwiat01,cia08,cia09}.
The state outgoing the two crystals may be written as
\begin{align}
| \Phi \rangle & = \frac{1}{\sqrt{2}} \int\!\!\! d\theta\,
d\omega_{s}\,d\omega_{p} f(\omega_{p},\omega_{s},\theta)
A(\omega_{p}) \nonumber \\
& [e^{\imath k_{p}^{o}(\omega_{p})L} e^{\imath \phi(\theta)
+\imath \phi'(\theta')} |H,\theta,\omega_{s}\rangle
|H,\theta',\omega_{p}-\omega_{s}\rangle +\nonumber \\ &
+ e^{\imath k_{2\parallel}(\omega_{p},\omega_s,\theta) L}
|V,\theta,\omega_{s}\rangle |V,\theta',\omega_{p}-\omega_{s}
\rangle ]\;,
\end{align}
where $L$ is the crystals length.
The complex amplitude spectrum of the
pump laser is $A(\omega_{p})$ whereas the downconverted photons are
generated with the two photons spectral and angular amplitude
$f(\omega_{p},\omega_{s},\theta)$, as defined in \cite{cia09}.  We call
$\overline{\omega_{p}}=\omega_{p}^0 + \omega_{p}$ and
$\overline{\omega_{s}}=\omega_{p}^0/2 + \omega_{s}$ the frequencies of
the pump laser and of the signal, where $\omega_{p}$ and $\omega_{s}$
are the shift from the central frequencies $\omega_{p}^0$ and
$\omega_{p}^{0}/2$. Likewise signal and idler generation angles are
respectively $\overline{\theta}=\theta^0+\theta$ and $\overline{\theta
'}=-\theta^0 +\theta '$, with $\theta^0$ the central angle and
$\theta'= -\theta +\gamma \omega_s + \gamma' \omega_p$. Within the
spectral width of our pump the dependence on $\omega_p$ is negligible 
and thus we have
$\theta'\simeq -\theta + \gamma \omega_s$, with
$\gamma= \partial \theta '/\partial \omega_s$.
\par
The phase term
$k_{p}^{o}(\omega_{p})L$ is due to the pump traversing the first crystal
before it generates photons in the second one, whereas the term
$k_{2\parallel}
(\omega_p,\omega_s,\theta)L$ appears because the photons generated in the 
first crystal must traverse the second one. The perpendicular part 
$k_{2\perp}$ disappears for conservation of the transverse momentum, as 
it is guaranteed by the large pump spot on the crystals ($\sim 1.5mm$).
The other phase terms are common and are grouped out.
The phase-shifts $\phi(\theta)$ and $\phi' (\theta')$ are introduced by
a spatial light modulator (SLM) respectively on the signal and on the
idler and depend on the generation angles $\bar\theta$ and $\bar\theta'$.
These will discussed in details in the following. It can be shown
numerically that $f(\omega_{p},\omega_{s},\theta)\approx
f(0,\omega_{s},\theta)\equiv f(\omega_{s},\theta)$ for crystals length
$L\lesssim 1 mm$.  Upon expanding all the contributions to the optical
paths to first order and after some algebra we arrive at
\begin{align}
| \Phi \rangle &=\frac{1}{\sqrt{2}} \int_{-\Delta \theta /2}^{\Delta
\theta /2}\!\!  d\theta \int_{\omega_{s1}(\theta)}^{\omega_{s2}(\theta)}
d\omega_{s} \int d\omega_{p} f(\omega_{s},\theta) A(\omega_{p})
\cdot\nonumber \\ &[|H,\theta,\omega_{s}\rangle |H,\theta',\omega_{p}
-\omega_{s} \rangle + \nonumber\\ &+e^{\imath \varphi(\omega_{p},\omega_{s},
\theta)}|V,\theta,\omega_{s} \rangle |V,\theta',\omega_{p}-\omega_{s}
\rangle ]
\label{Eq_Phi}
\end{align}
where $\Delta \theta$ is the angular acceptance and
$\omega_{s1,2}=1/\gamma(\theta\mp\Delta\theta/2)$ are the integration
limits for $\omega_s$, as determined by $\Delta \theta$ and $\gamma$.
The phase function between the $H$ and the $V$ component is given by
$$\varphi (\omega_{p},\omega_{s},\theta)= \phi_{0}+\alpha L
\omega_{p}+ \beta L \omega_{s} - \delta L \theta -\phi(\theta) -
\phi'(\theta')\:,$$ where $\phi_0$ includes all the zero-order terms of the
expansion. The phase term $\alpha L \omega_p$ accounts for the delay
time between horizontal and vertical downconverted photons.  The two
subsequent terms rise for conservation of the transverse momentum.  The
term $\beta L \omega_s$ may be understood by considering the signal at
the fixed angle $\overline{\theta}$: for different $\omega_s$
the idler sweeps different $\overline{\theta'}$ and this means
different optical path.  Likewise fixing
$\omega_s$, a positive variation of $\theta$ correspond to a negative
variation of $\theta'$ and this introduces an optical path dependent on
$\theta$, i.e. the phase shift $\delta L \theta$.  The delay time between
the photons may be compensated upon the introduction of a proper
combination of birefringent crystals on the pump path, as already
demonstrated in \cite{cia08} (see Fig. \ref{2D_generation}).
Let us now focus attention on the action of the SLM, i.e. on the
phase function $\phi(\theta)$ and $\phi'(\theta')$.  At first, 
since $\theta'\simeq -\theta + \gamma \omega_s$ we note that the 
choice $$\phi'(\theta')= \beta L \theta'/\gamma + \epsilon' \quad 
\phi(\theta)=-\beta/\gamma L \theta + \epsilon\,,$$ with 
$\epsilon'+\epsilon=\phi_0$ allows one to compensate all the remaining 
phase terms in $\phi (\omega_{p},\omega_{s},\theta)$ and to achieve 
purification of the state. In this way, we may generate polarization 
entangled states as in Eq. (\ref{Eq_Phi}) with $\phi(\omega_{p},\omega_{s},
\theta)=0$. Experimentally we obtain a visibility of about $\sim 0.9$
starting from $\sim 0.42$.
\section{Two-photon multipartite entanglement}
\begin{figure}[h!]
\includegraphics[angle=270,width=0.94\columnwidth]{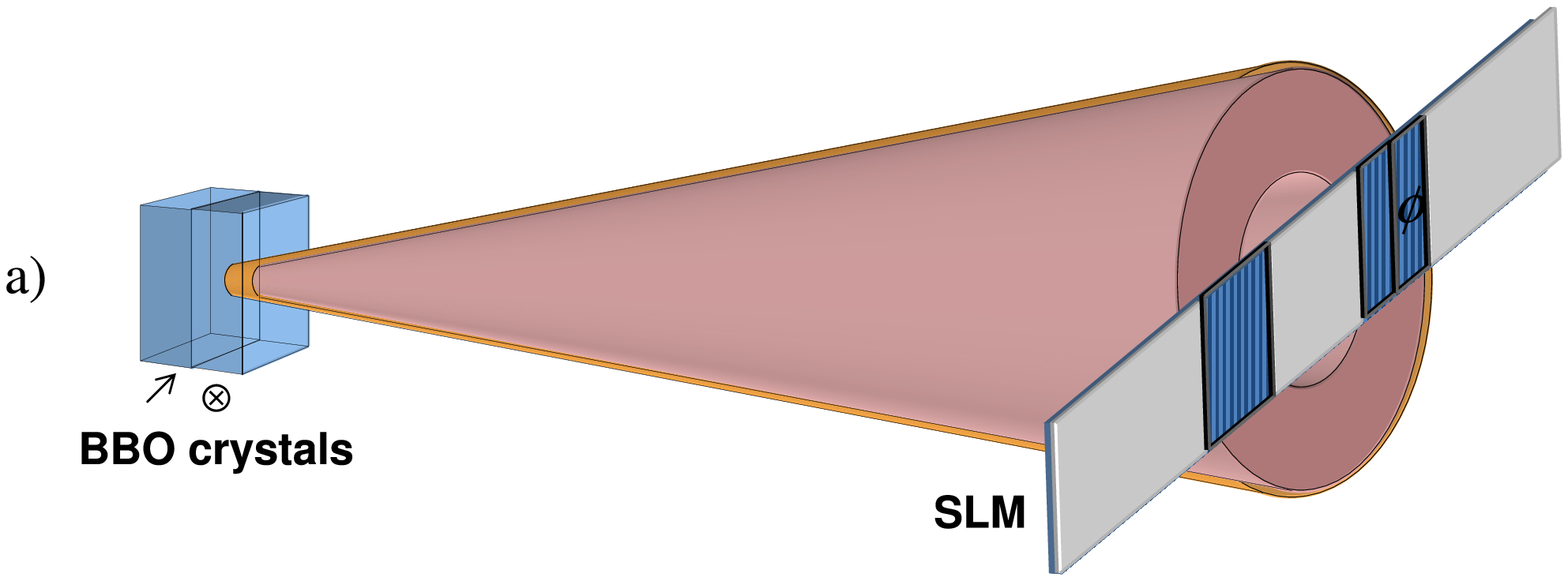}
\includegraphics[angle=270,width=0.94\columnwidth,clip=]{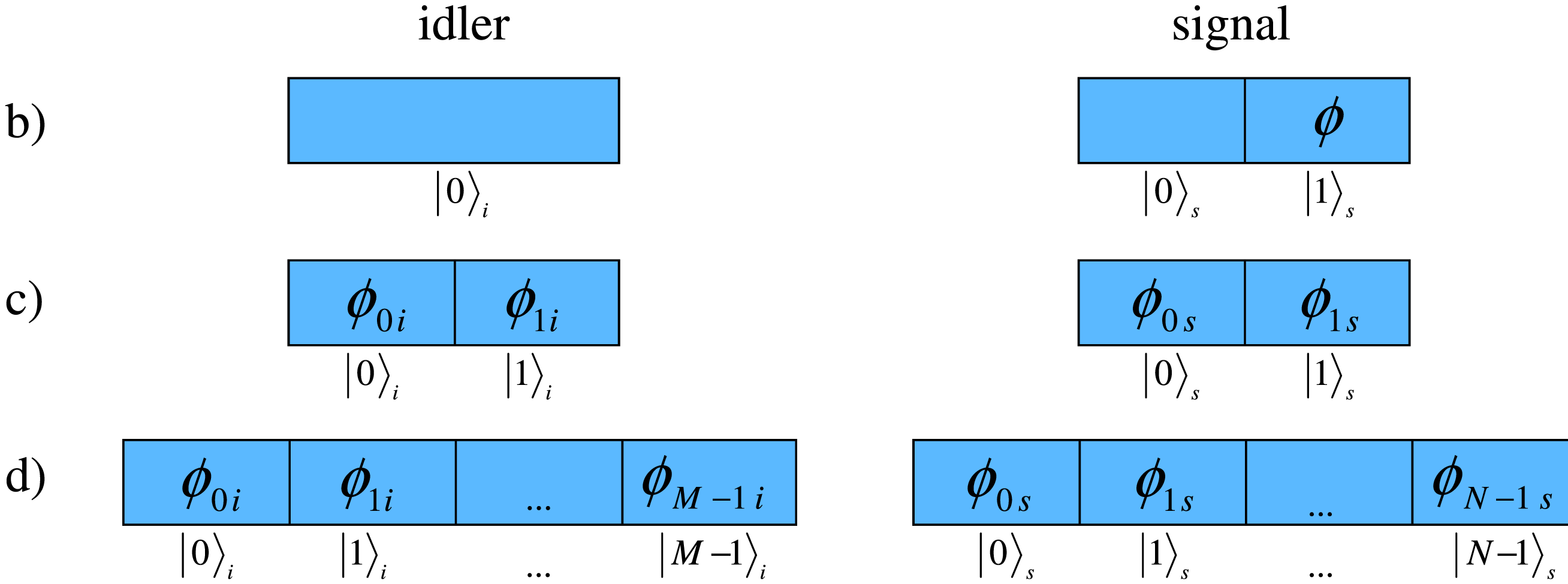}
\caption{Generation of multipartite entangled/cluster states by the
use of SLM. The overall output from SPDC is divided in spatial sections,
and a different phase may be imposed to each portion in a programmable
way, corresponding to the application of controlled phase-gates.
Each spatial section includes several pixels. In (b) we report the
momentum state in our experimental implementation $ \ket{\Psi}=
\frac{1}{\sqrt{2}}\ket{0}_i(\ket{0}_s+\ket{1}_s)$; in (c) the state
configuration to achieve $\ket{\Psi}=\frac{1}{2}(\ket{0}_i
+\ket{1}_i)(\ket{0}_s+\ket{1}_s)$ (which may become
$\ket{\Psi}=\frac{1}{\sqrt{2}}(\ket{0}\ket{1}+
\ket{1}\ket{0})$ upon the use of a spectral filter, bandpass of $10
nm$). In (d) the generic configuration leading to the momentum
state $\ket{\Psi}=\ket{s}\otimes\ket{i}$.}
\label{ClG}
\end{figure}

In order to generate multipartite entangled states, and in particular 
cluster states, purification is just the first step. Here
we suggest a new technique based on the use of the SLM.
We consider the signal and the idler beams divided in N and M
subdivisions (see Fig. \ref{ClG}(d)), which individuate different momentum
qudits, and write the signal and idler momentum state as
$$|s\rangle=\sum_n^N a_n|n\rangle_s\,, \quad |i\rangle=\sum_m^M
a_m|m\rangle_i$$
with $n=0,1,...,N-1$ and $m=0,1,...,M-1$.  The total momentum state is
$|\Psi\rangle=|s\rangle \otimes |i\rangle$.  This is not an entangled
state in the momentum since for a certain signal angle $\theta$, the
idler sweeps a wide interval of $\theta'$, actually covering all the
angular acceptance $\Delta \theta$ due to the broad down
conversion spectrum.
The global state is thus given by $|\Phi\rangle \otimes |\Psi\rangle$,
where polarization provides two qubits, and the rest of information is 
encoded onto the momentum degrees of freedom \cite{add1,add2,add3}.  
\par
The action of the SLM corresponds to impose a phase shift only
on the horizontal component of polarization, leaving the vertical part
undisturbed. We exploit this property to add a different constant phase,
besides the purification ones $\phi(\theta)$ and $\phi'(\theta')$,
for each
portion of signal and idler. This corresponds to the action of a set
of controlled phase-gates $C_{\boldsymbol \phi}$,
$\boldsymbol\phi=\{\phi_{0i},...,\phi_{M-1i}, \phi_{0s},...,\phi_{N-1s}\}$
to the state $|\Phi\rangle \otimes |\Psi\rangle$. Using a suitable
number of sectors (power of two) one may generate multiqubit entangled 
states of the form $|\Xi\rangle=C_{\boldsymbol \phi}|\Phi\rangle
\otimes |\Psi\rangle$.
\par
The simplest
example, which we implemented experimentally, is obtained using $M=1$
and $N=2$, i.e. by dividing the signal beams in two parts exploiting the
SLM to apply a phase $\phi$ to only one of them (see Fig. \ref{ClG}(a)
and (b)). This leads to the
generation of a two-photons three-qubit entangled state of the form
$\frac{1}{2}[ |000\rangle+|110\rangle+e^{\imath
\phi}|001\rangle+|111\rangle]$ where, for the first two qubits,
$|0\rangle\equiv |H\rangle$ and $|1\rangle \equiv |V\rangle$ whereas the
third qubit is the signal momentum. For $\phi=\pi$ one obtains a
two-photons three-qubit linear cluster state
$$|C_3\rangle=\frac{1}{\sqrt{2}}[|\Phi^{+}\rangle |0\rangle
-|\Phi^{-}\rangle |1\rangle]\,,$$ where $|\Phi^{\pm}\rangle$ are standard
Bell states. In order to highlight the power of our method let us
consider another example, with four qubits: for $M=N=2$ (see Fig. \ref{ClG}(c))
and applying $\phi_{0s}=-\phi_{0i}$, $\phi_{1i}=\pi - \phi_{1s}$
we achieve the four-qubit entangled
state 
\begin{align}
|\Xi_4\rangle=& \frac{1}{2}\big[
|\Phi^{+}\rangle
|00\rangle-|\Phi^{-}\rangle |11\rangle\notag \\ &+ |\Delta^{+}(\phi_{t})\rangle
|01\rangle - |\Delta^{-}(\phi_{t})\rangle |10\rangle\big]\,,\notag
\end{align}
where
$\phi_{t}=\phi_{0i}+\phi_{1s}$ and
$|\Delta^{\pm}(\phi_{t})\rangle=\frac{1}{\sqrt{2}} [|00\rangle \pm e^{\mp\imath
\phi_{t}}|11\rangle]$.  We foresee that using a narrower spectral filter
for the downconverted photons it is possible to select different regions
of the angular distribution in a way that allow us to engineer
entanglement also for the momentum degrees of freedom. Using a $10nm$ bandpass
filter and coupling $\Delta \theta \simeq 1.6 mrad$ on the momentum
channels $\ket{n}_s,\ket{m}_i$, for $N,M=2$,
we would have $\ket{\Psi}=\frac{1}{\sqrt{2}}(\ket{0}\ket{1}+
\ket{1}\ket{0})$. In such a way the total state would be the two-photon
four-qubit cluster state reported in \cite{chen07,val07}.
\subsection{Experimentals}
\begin{figure}[h!]
\begin{center}
\includegraphics[angle=270,width=0.95\columnwidth]{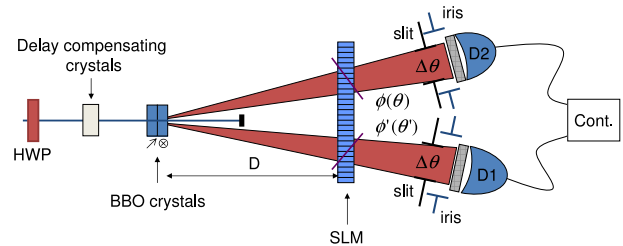}
\caption{Experimental setup. Polarization entangled states are generated
by downconversion in type-I BBO crystals pumped @$405nm$ by a
Newport LQC$405-40$P CW laser diode. Then, a spatial section of
the output cones passes through the SLM which (at the same time)
i) provides
purification of polarization entanglement and ii) introduces
a position- (i.e. generation angle) dependent phase-shift between
the two polarizations.
After SLM there are (on both paths) a slit, an iris, two longpass
filters (cut-on wavelength $= 715nm$), a coupler with an $1/e^2$
output beam diameter of $7.14mm$ and a multimode
optical fiber that directs the photons to the detector. The detectors
are home-made single-photon counting modules ($D1,D2$), based on an
avalanche photodiode operated in Geiger mode with passive quenching.
For tomographic reconstruction we insert a quarter-wave plate, 
a half-wave plate and a polarizer and for the optimization of the 
phase functions only the polarizers.
}\label{2D_generation}
\end{center}
\end{figure}

The experimental setup is shown in
Fig. \ref{2D_generation}.  The pump
derives from a $405nm$ CW laser diode (Newport LQC$405-40$P).  After an
half wave plate (HWP) that rotates its polarization in order to balance
the generated state, the pump passes through two BBO crystals that
compensate the delay time $\alpha L$ between $\ket{H}$ and $\ket{V}$
generated photons. Two BBO crystals, each cut for type-I down
conversion, stacked back-to-back and oriented with their optical axes
aligned in perpendicular planes, are used to generate the polarization
entangled state.  As shown in Fig. \ref{ClG}(a) a portion of the output
cones passes through the spatial light modulator (SLM), which is a
crystal liquid phase mask ($64\times10mm$) divided in $640$ horizontal
pixels, each wide $d=100\mu m$ and with the liquid crystal $10\mu m$ deep.
The SLM is set at a distance $D=500 mm$ from the two generating crystals.
Driven by a voltage the liquid crystal orientation in correspondence of
a certain pixel changes. The photons with an horizontal polarization
feel an extraordinary index of refraction depending on the orientation,
and this introduce a phase-shift between the two polarizations.  Since
each pixel is driven independently we can introduce a phase function
dependent on the position on the SLM, i.e. on the generation angle
$\theta$ and $\theta'$.  
\par
It is worth noting that the SLM also replaces
the birefringent plate used for the optimal generation of photon pairs
\cite{Kwiat01}. After SLM, on the signal and the idler paths, there are
a slit, an iris, two longpass filters (cut-on wavelength $= 715nm$), a
coupler with an $1/e^2$ output beam diameter of $7.14mm$ and a multimode
optical fiber that directs the photons to the detector. The detectors
are home-made single-photon counting modules ($D1,D2$), based on an
avalanche photodiode operated in Geiger mode with passive quenching.
For the tomographic reconstruction we insert, on both paths after 
the iris, a quarter-wave plate, a half-wave plate and a polarizer.
For the optimization of the phase functions (see below) we
insert only the polarizers.
\par
Our experimental setup allows us to collect the downconverted photons
within a wide spectrum and angular distribution. In order to underline
this fact the pump power on the crystals has been intentionally left
very low ($2.5mW$) by using an amplitude modulator. To collect as many
photons as possible we make the imaging of the pump spot on the crystals
($\simeq 1.5mm$) into the optical fibers core (diameter of $62.5\mu m$)
using the coupler lenses. Setting the slits at $4mm$ ($\Delta \theta
\simeq 6.5 mrad$) and the iris with a diameter of $9mm$ we collect up to
$100$ coincidence couns per second. It is worthwhile
to note that such an angular acceptance $\Delta \theta$ acts as a
$100nm$ bandpass spectral filter for the down converted photons. In order
to purify the state we insert the phase functions 
$$\phi(x)=a_2(x-x_{c2}) +b_2 \qquad \phi'(x)=a_1(x-x_{c1})+ b_1\,,$$
where $x$ is the pixel number,
$x-x_{c2}=\frac{D}{d}\theta$ and $x-x_{c1}=\frac{D}{d}\theta'$,
$x_{c1}$ and $x_{c2}$ are the central pixels on idler and signal,
i.e. the pixels corresponding to the central angles $\overline{\theta'}=-\theta^{0}$
and $\overline{\theta}=\theta^{0}$. The values of the parameters $a_1,
b_1, a_2$ and $b_2$ has been optimized upon inserting two polarizers 
set at $\alpha_1=45^{\circ}$ and $\alpha_2= -45^{\circ}$ 
in front of the couplers and then searching for the minima in the 
coincidence counts, corresponding to the values of $b_{1,2}$ compensating 
the constant phase difference $\phi_0$ and $a_{1,2}$ removing the 
angular dependence on $\theta$ and on $\theta'$, and in turn on
$\omega_s$. For our configuration 
we have $a_1=-a_2=\beta L d/\gamma 
D\simeq - 0.05$, $b_1+b_2=\phi_0$.
\begin{figure}[h]
\includegraphics[angle=270,width=0.8\columnwidth]{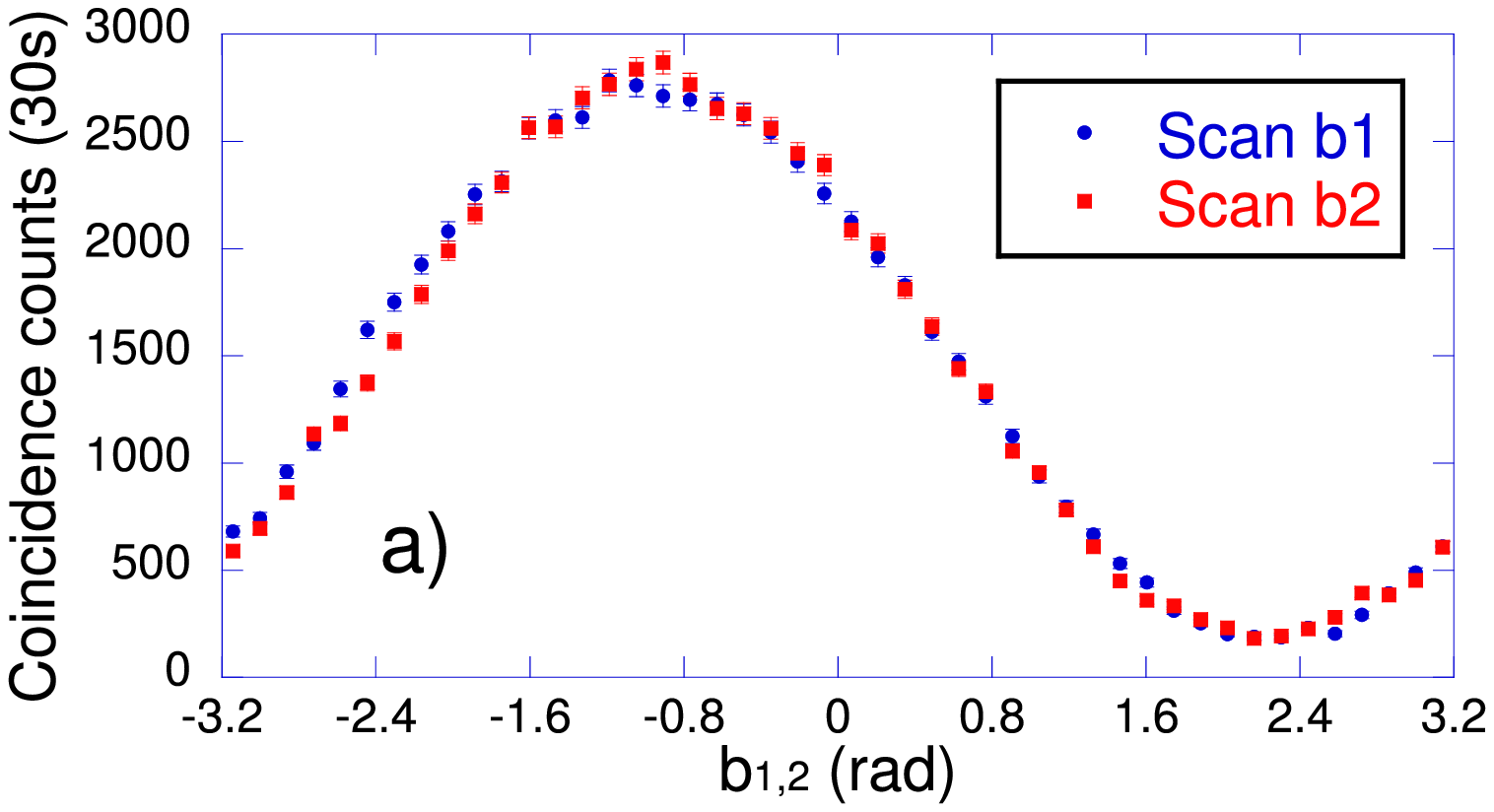}
\includegraphics[angle=270,width=0.8\columnwidth]{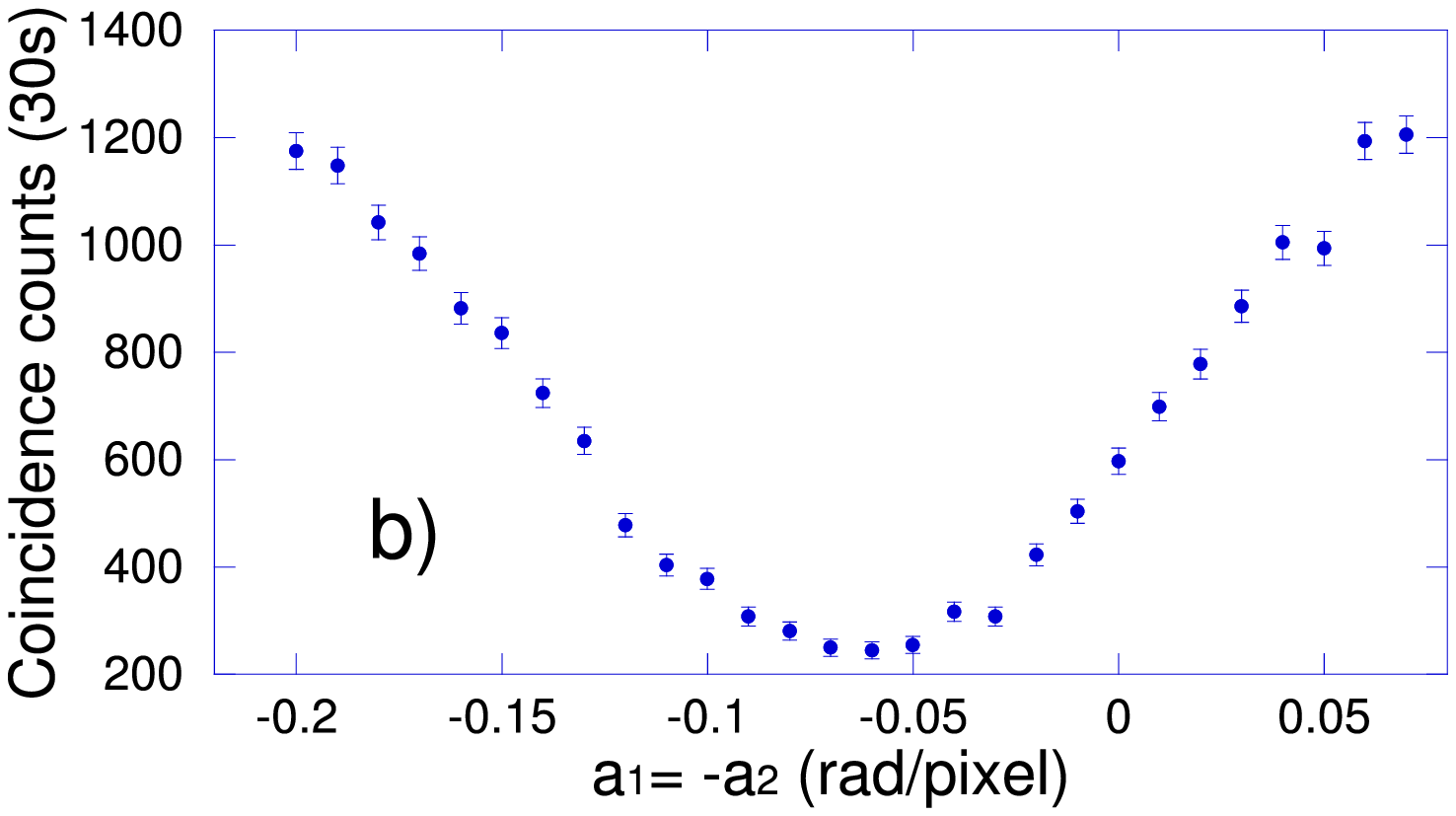}
\caption{Coincidence counts on a time window of $30s$ with 
the polarizers in front of the couplers set at $45^{\circ}$ and 
$-45^{\circ}$. (a) Coincidences as a function of $b_{1,2}$ (blue/red) 
with $b_{2,1}=0$ and optimal $a_{1,2}$; (b) Coincidences as a function
of $a_1=-a_2$ with optimal $b_1$ and $b_2=0$. 
\label{scan_a_and_b} }
\end{figure}
\par
In Fig.  \ref{scan_a_and_b}(a)we report the coincidence counts on a time 
window equal to $30s$ as a function  of $b_{1}$ ($b_2$) (blue/red) with 
$b_2=0$ ($b_1=0$) and with $a_{1,2}$ set to their optimal values.  In Fig. 
\ref{scan_a_and_b}(b) we report the coincidence counts on a time window 
of $30s$ as a function of $a_1=-a_2$ with $b_2=0$ and $b_1=\phi_0$. The 
agreement with the theoretical model is excellent. In turn, the purification 
of the state works as follows: starting from a visibility equal to $0.423\pm
0.016$ we achieve $0.616 \pm 0.012$ after the delay compensation with the 
crystals and $0.886 \pm 0.012$ after the spatial compensation with the SLM. 
Finally, by closing the iris at the same width of the slits we obtain $0.899 
\pm 0.008$. Actually, we verified experimentally that variations of the 
phase in the azimuthal direction have only a minor effect. The residual lack 
of visibility is in turn due to the low spatial coherence of the pump, which 
is spatially multimode. 
\subsection{State reconstruction}
Upon properly programming the SLM, i.e. by
setting $M=1$, $N=2$, and $\phi=\pi$ as in Fig. \ref{ClG}(b), our
scheme may be set to generate, in ideal conditions, the cluster
state $|C_3\rangle$.
In order to characterize the output state, denoted by
$R_3$, and to check the effects of the decoherence processes,  we have
performed state reconstruction by (polarization) quantum tomography
\cite{mlik00,jam01}.  
The experimental procedure goes as follows: upon measuring
a set of independent two-qubit projectors $P_\mu= |\psi_\mu\rangle
\langle\psi_\mu |$ $(\mu=1,...,16)$ corresponding
to different combinations of polarizers and phase-shifters, the
density matrix may be reconstructed as $\varrho=\sum_\mu p_\mu\,
\Gamma_\mu$ where $p_\mu = \hbox{Tr}[\varrho \, P_\mu]$ are the
probabilities of getting a count when measuring $P_\mu$ and
$\Gamma_\mu$ the corresponding dual basis, {\em i.e.} the set of
operators satisfying $\hbox{Tr}[P_\mu\,\Gamma_\nu] =
\delta_{\mu\nu}$ \cite{dar01}.
Of course in the experimental reconstruction the probabilities
$p_\mu$ are substituted by their
experimental samples {\em i.e.} the frequencies of counts obtained
when measuring $P_\mu$. In order to minimize the effects of
fluctuations and avoid non physical results we use
maximum-likelihood reconstruction of two-qubit states
\cite{mlik00,jam01}. 
\par
\begin{figure}[h!]
\includegraphics[width=0.45\columnwidth]{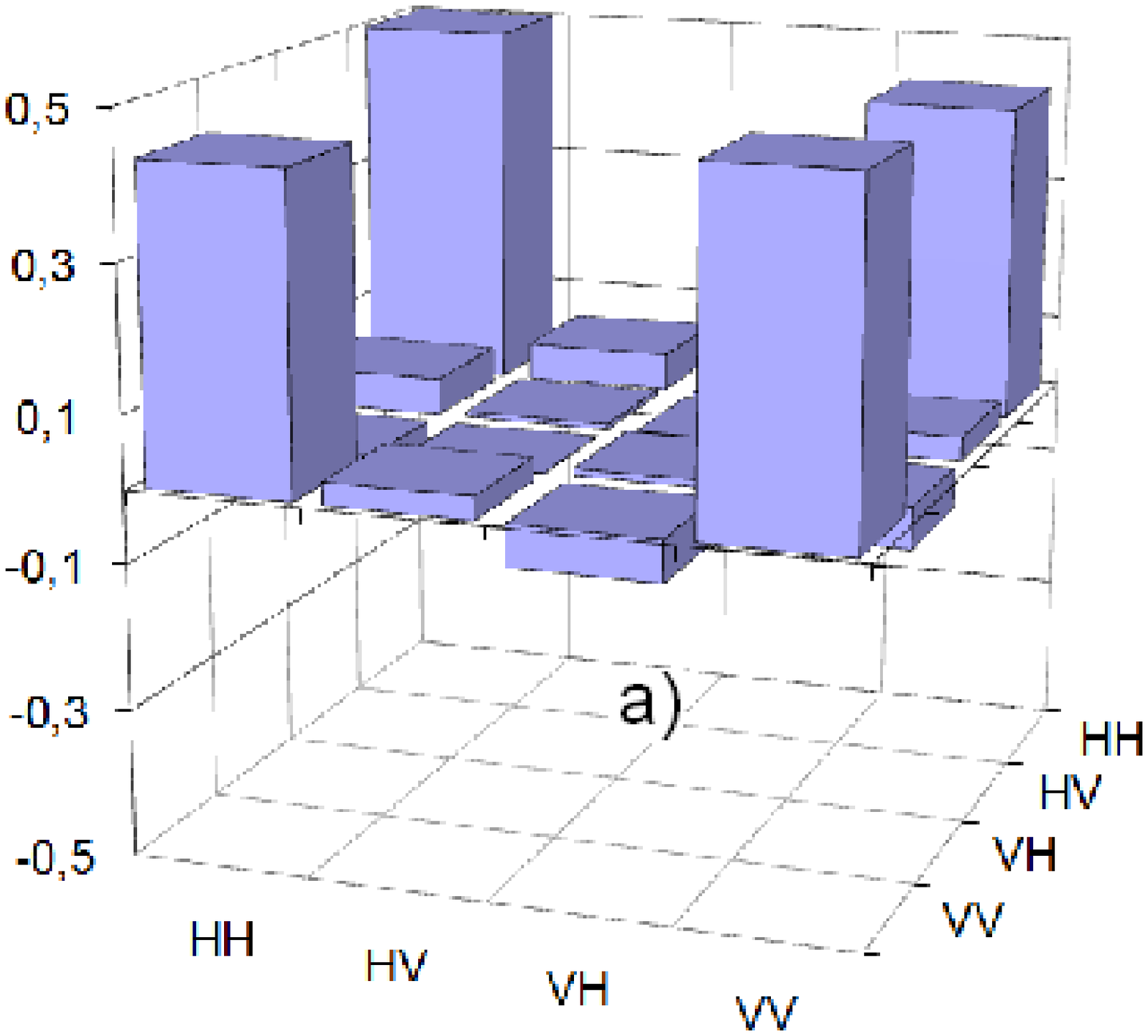}
\includegraphics[width=0.45\columnwidth]{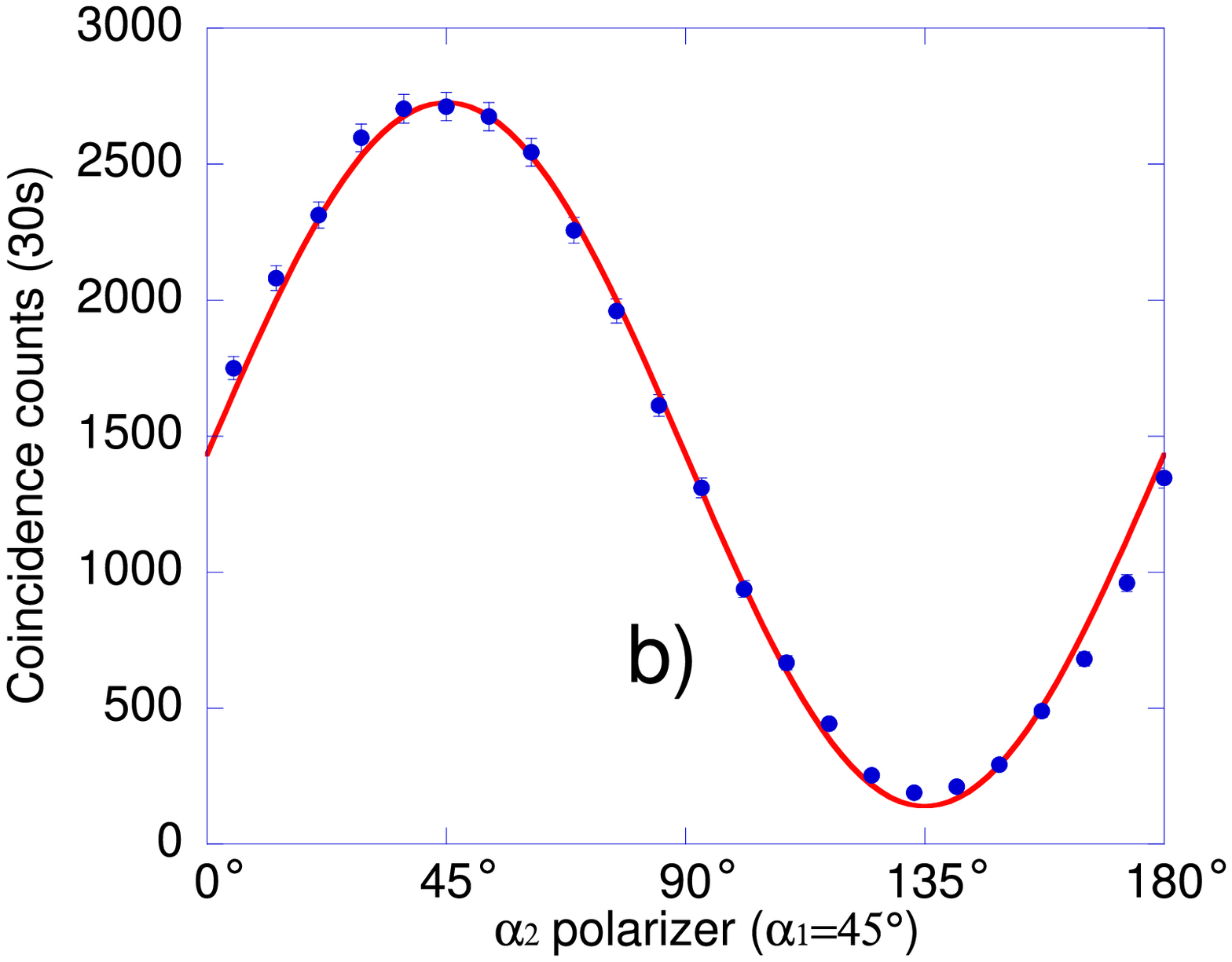}
\includegraphics[width=0.45\columnwidth]{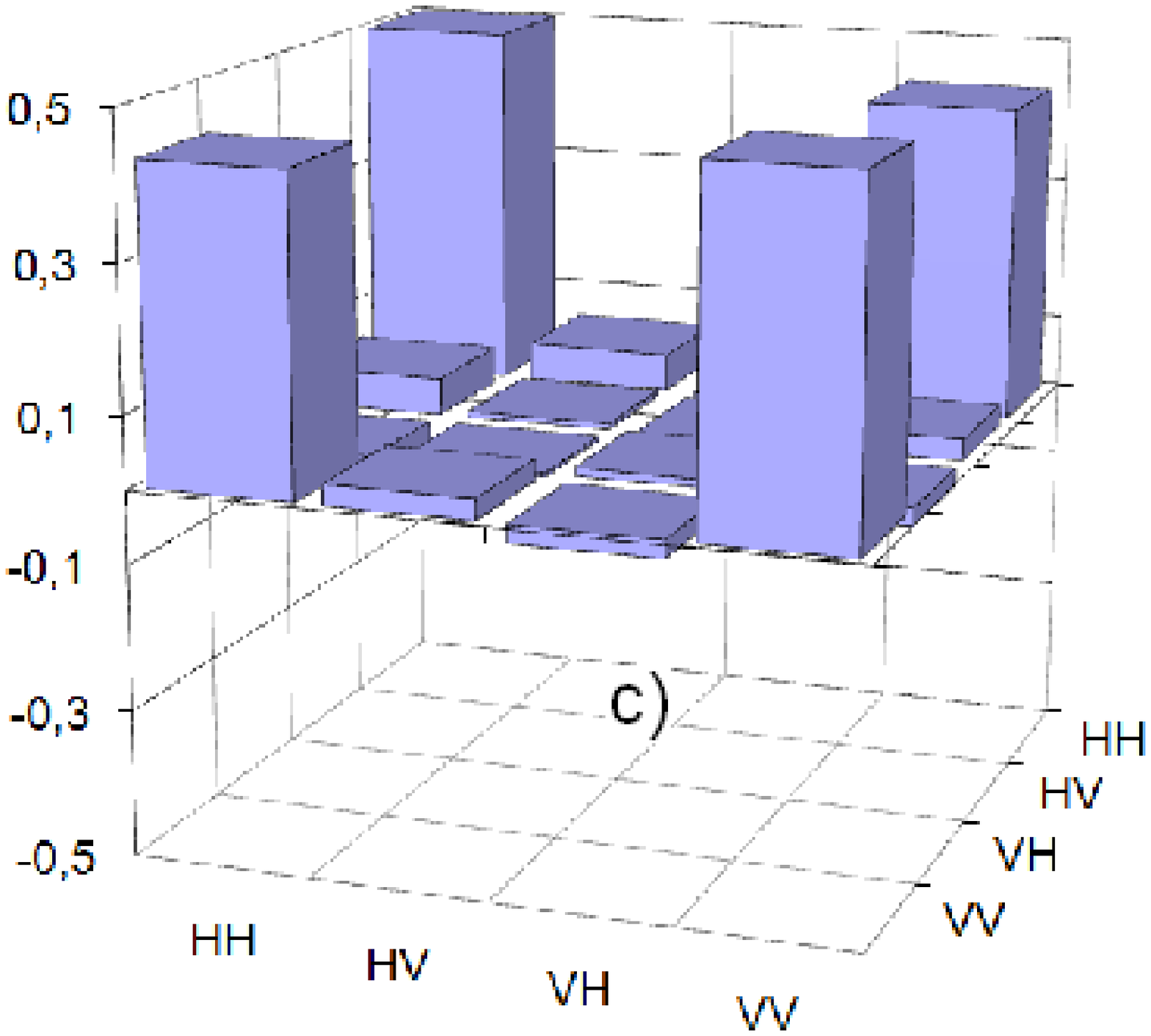}
\includegraphics[width=0.45\columnwidth]{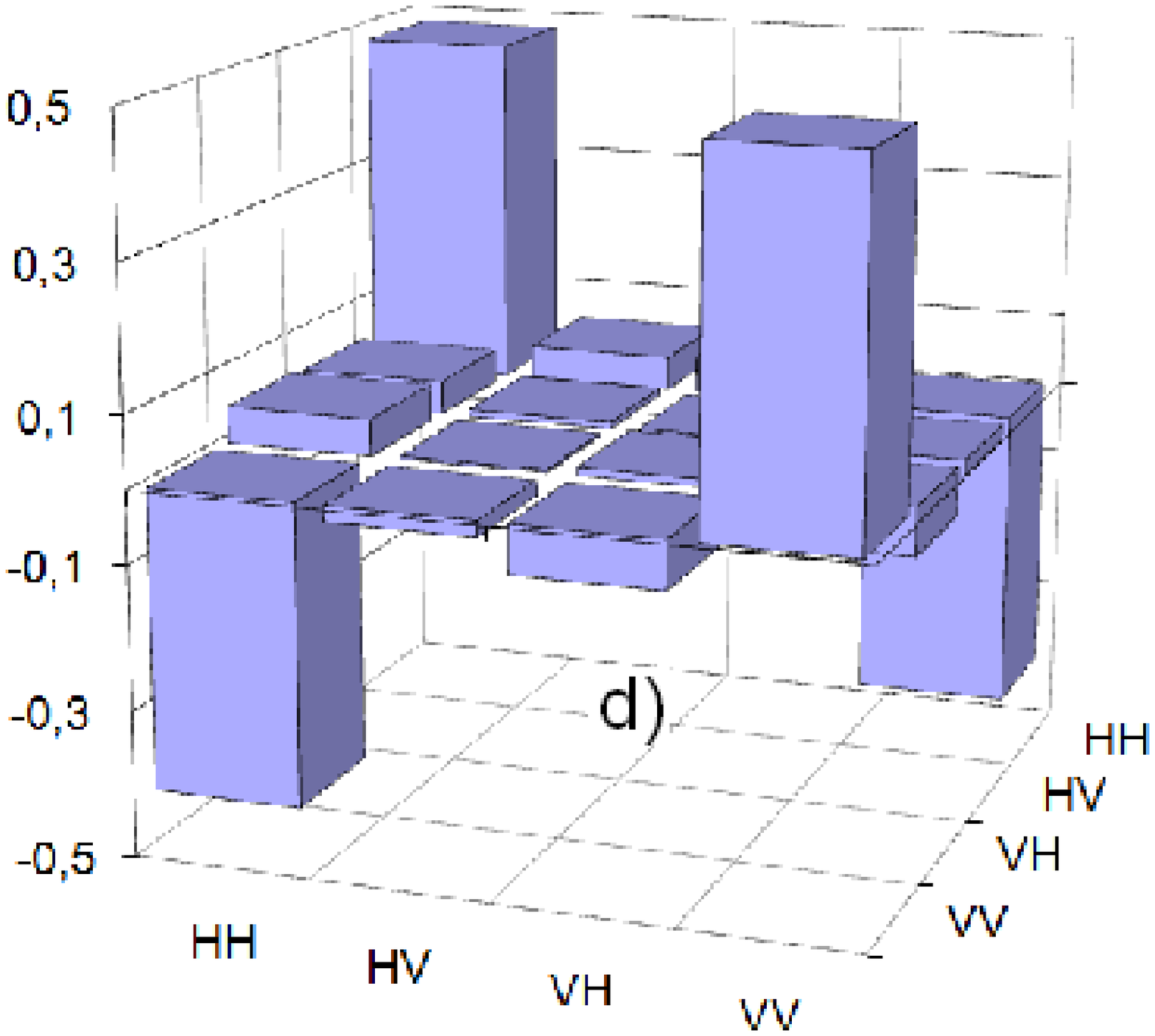}
\caption{Characterization of the output state. In (a) we report the
tomographic reconstruction (real part) of the global purified 
polarization entangled state prior the action of the phase-gate, 
whereas in (b) we show the corresponding visibility curve and the 
fit with the curve $\cos^2(\alpha_2-45^{\circ})$ (solid line). 
In (c) and (d) we report the tomographic reconstructions (real part) 
of the reduced states
$\varrho_0$ and $\varrho_1$.}\label{qt}
\end{figure}

At first we reconstruct the purified state prior
the action of the SLM phase-gate, i.e. without addressing the momentum
qubit. Then, we reconstruct the two reduced states $\varrho_j= \frac{1}{p_j}
\hbox{Tr}_3[ \ket{j}_{s}{}_s\bra{j}\,R_3 ]$ obtained by measuring the
momentum qubit after the phase-gate. This is obtained by moving the slit
on the signal to select the corresponding portion of the beam.  Results
are summarized in Fig. \ref{qt}. 
\par
As it is apparent from the plots our
scheme provide a reliable generation of the target states.
Fidelity of the purified polarization state is about $F\simeq 0.90 \pm
0.01$,
whereas fidelities of the conditional states $F_0= \bra{\Phi^+} \varrho_0
\ket{\Phi^+}$ and $F_1= \bra{\Phi^-} \varrho_1 \ket{\Phi^-}$ are given
by $F_0=0.92\pm 0.01$ and $F_1=0.90\pm 0.01$ respectively. In order to 
achieve this precision we have employed a long acquisition time 
($\sim 60 s$) thus also demonstrating the overall stability of 
our scheme.
We also report the visibility of the state prior the action of the SLM 
phase-gate, which confirms the entanglement purification process
\cite{EEE10}.
\section{Conclusions}
We have suggested and implemented a novel scheme
for the generation of two-photon multipartite entangled states. In our
device a programmable spatial light modulator acts on different
spatial sections of the overall downconversion output and provides
polarization entanglement purification as well as
arbitrary phase-gates between polarization and momentum qubits.
It should be mentioned that also measurements on the momentum
qubits benefits from our configuration. In fact, addressing
momentum is equivalent to select portions of the signal (idler)
beam and then make them interacting, say by a beam splitter and other
linear optical elements,
to perform arbitrary momentum measurements. 
In our scheme this may be implemented in a compact form since
the portions of the beam are quite close each other, and we may 
work with beam splitter at non normal incidence.
Overall, our scheme represents an effective, low cost, source of
two-photon multiqubit/qudit entanglement. We foresee applications in
one-way quantum computation and quantum error correction.
\acknowledgments
The authors thank I. Boscolo for encouragement and support.

\end{document}